# Immunoexpression of Nm23: correlation with traditional prognostic markers in breast carcinoma.

Esther Durán, and Riánsares Arriazu[*].

*Histology Laboratory, Institute of Applied Molecular Medicine, Department of Basic Medical Sciences, School of Medicine, CEU-San Pablo University, Madrid, Spain.*

**ABSTRACT-** Fifty cases of breast carcinoma were investigated for Nm23, LPA1, Ki67 and p53 antigen expression, using immunohistochemical techniques. Correlation between markers was studied. Nm23 showed an inverse correlation with LPA1 and Ki67, and positive with p53. Nm23/LPA1 presented significant (p<0.001) negative correlation with p53, but not difference were observed with Ki67. It can be concluded that: 1) Nm23 and LPA1 tend to be inversely correlated; 2) Nm23 correlates positively with p53, regardless of tumour type and patient's hormonal status; 3) Nm23 shows a reverse trend in comparison with Ki67. However, more studies are needed to determinate if hormone receptors play some role in the inverse relationship between Nm23-LPA1 and to know the tumour type implication in these relationships.

## INTRODUCTION

Breast cancer is a heterogeneous disease, comprising multiple entities associated with distinctive histological and biological features, clinical presentations and behaviours and responses to therapy [1;2]. Tumor cell invasion is a complex process and the mechanisms controlling the transition from ductal carcinomas *in situ* to invasive ductal carcinoma still remain unclear, despite research in recent years [3].

Various biological makers known to be indicators of prognosis in breast cancer include growth factor receptors, estrogen receptors, p53, and proliferation indices like Ki67. Majority of breast cancer patients succumb to metastatic disease.

The molecular basis of the metastatic disease is not known, but activation or inactivation of multiple genes is involved in the various steps of tumour progression [4;5].

The first Nm23 gene was discovered by Steeg et al. in 1988 [6] on the basis of its reduced expression in the highly metastatic murine myeloma cell lines, as compared to their nonmetastatic counterparts [6;7].

Similar trends were identified in other model systems. Low Nm23-H1 expression in human tumours often correlated with poor patient survival although it is not considered to be an independent prognostic factor [8].

Another growing target of interest is LPA1 (also known as EDG2), one of the receptors for LPA (lysophosphatidic acid). LPA1 is involved in a variety of physiological functions such as cellular proliferation, differentiation, prevention of apoptosis, migration, invasion, regulation of actin cytoskeleton, modulation of cell shape, and survival [9;10].

The aim of this study was to investigate: 1) the immunohistochemical expression of Nm23 and LPA1 in breast carcinomas; 2) correlation between expressions of Nm23 with LPA1 and with other traditional markers in breast cancer diagnosis; and 3) correlation between expressions of Nm23/LPA1 and with other clinicopathology variables.

## MATERIALS AND METHODS

Formalin fixed paraffin-embedded blocks of breast lesions were retrieved from the archives of the Service of Pathology, Hospital Madrid Montepríncipe, Madrid, Spain, where classified the carcinomas in histological type and grade. There were 50 samples in total distributed into two groups as: invasive ductal carcinoma (grade 1 (n = 4), 2 (n = 9) or 3 (n = 21)), and invasive lobular carcinoma (mixed (tubulolobular) carcinoma grade 2 (n = 1), lobular carcinoma (n = 12), and pleomorphic lobular carcinoma grade 3 (n = 3)).
Blocks of breast lesions were serially sectioned at five µm-thicknesses and stained with hematoxylin-eosine or used for immunohistochemical techniques.

[*] Correspondence to: R. Arriazu, Department of Basic Medical Sciences, School of Medicine, San Pablo-CEU University
e-mail: arriazun@ceu.es

**Table 1**. Primary antibodies used for immunohistochemistry.

| Antigen | Concentration | Supplier | Source |
|---|---|---|---|
| Nm23-H1 | 1:100 | NOVUS BIOLOGICALS Litttleton. USA | Monoclonal |
| ER (estrogen receptor) | Ready to use | DAKO Carpinteria, California. USA | |
| PR (progesterone receptor) | Ready to use | DAKO Carpinteria, California. USA | |
| Ki67 | Ready to use | DAKO Carpinteria, California. USA | |
| p53 | Ready to use | DAKO Carpinteria, California. USA | |
| LPA1(EDG2) | 1:100 | NOVUS BIOLOGICALS Litttleton. USA | Polyclonal |

**Immunohistochemical Methods**

Slices were deparaffinized and rehydrated tissue sections were treated for 30 min with hydrogen peroxide 0.3% in phosphate-buffered saline (PBS) pH 7.4, to block endogenous peroxidase, antigen unmasking was performed with pepsin (15 min, Sigma R2283). To minimize nonspecific binding, sections were incubated with serum blocking solution (Histostain® Bulk Kit, Invitrogen Corporation, Carlsbad, California) and subsequently incubated overnight with primary antibodies in a moist chamber at 4ºC. Information on primary antibodies is provided in Table 1.

The second day, immunohistochemistry was performed with standard procedures using Histostain® Bulk Kit (Invitrogen Corporation, Carlsbad, California). After immunoreactions, sections were counter-stained with Harris hematoxylin, dehydrated in ethanol, and mounted in a synthetic resin (Depex, Serva, Heidelberg, Germany). The specificity of the immunohistochemical procedures was checked by incubation of sections with nonimmune serum instead of the primary antibody.

**Clinical Methods**

ER, PR, Ki67, and p53 were evaluated by the pathologist visually, providing a semi-quantitative estimate of the percentage of positive cells.

**Image Analysis**

To establish the immunostaining percentage area of Nm23 and LPA1, ten digital images were acquired for each slide. To determinate the threshold values that discriminated the immunostaining from the counterstain and background, the red-green-blue (RGB) color mode values were converted to HSI values using MetaMorph software (Leica MMAF 1.4). The HSI system is thought to offer a much closer approximation to the behavior of human color vision than do untransformed RGB values [11].

We next defined the "negative-chromaticity subdomain" following the methodology described by Maximova et al. (2006) [12]. Finally, we defined a positive-chromaticity subdomain—a set of HSI threshold value ranges selected such that no pixel with the negative-control hue would confound the identification of positively stained areas.

**Statistical Analisis**

Statistical analysis was conducted using Pearson's correlation coefficient test. Differences were considered significant when $p < 0.001$. Missing data were removed in pairs. The software used was Excel 2010 version 14.0.6 (SP1) (Microsoft Corporatio. Redmond, Washington, USA).

## RESULTS

### Qualitative results

Nm23 expression was strongly detected in epithelial cytoplasm of normal acini. Epithelial cells that had lost their normal histological structure also showed Nm23 staining (figs. 1a,b). LPA1 expression was identified at epithelial cytoplasm as occurred with Nm23, but in this case, immunostaining was weaker (figs. 1c-d).

Immunoreactivity to p53 was identified in cytoplasm and nuclear epithelial cells from all samples studied (figs. 2a,b). Nuclear p53 immunostaining was also remarkable in the acini with histological changes (fig. 2b). Ki67 immunoreactivity was observed in the nuclei of epithelial cells of the acini. There were two types of Ki67 staining: homogeneous nuclear staining, and granular staining. The first one was more intensive than the second one (fig. 3).

ER and PR immunoexpression was found in the nuclei of epithelial cells from all samples studied (figs. 4a,b).

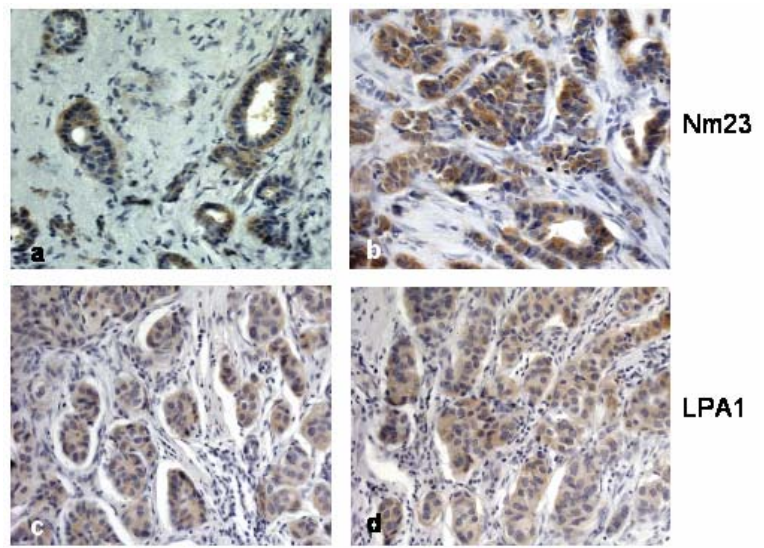

**Figure1.** Nm23 immunostaining (a,b) and LPA1 (c,d) x200. **a**: Acini with the cytoplasm immunostaining for Nm23. **b**: Tumour region intensely inmunostaining. **c**: Immunopositive acini for LPA1. **d**: Unstructured epithelium with LPA1 immunorexpression in the cytoplasm.

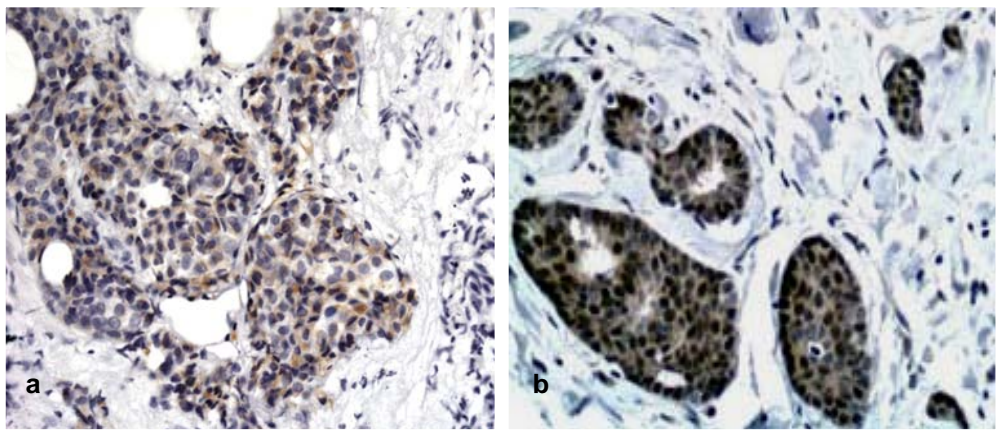

**Figure2.** p53 expression x200. **a**: Diffuse expression was observed in the cytoplasm of epithelial cells. **b**: Intense nuclei immunostaining.

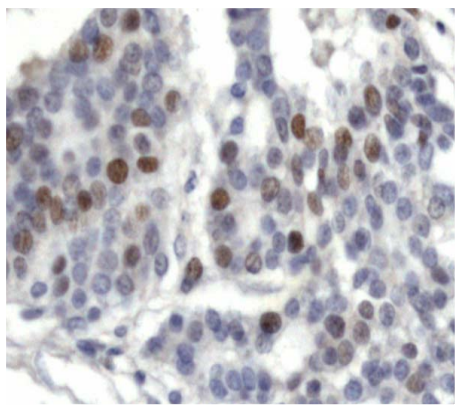

**Figure3**. Ki67 Inmunoxpresión x200. The image showed epithelial nuclear staining.

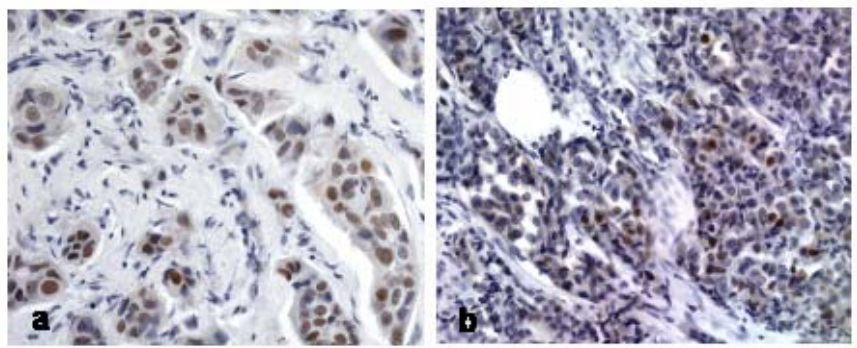

**Figure4**. Hormone Receptor nuclear immunostaining x200. **a**: Positive immunohistochemical reaction for ER. **b**: Positive immunoexpression for PR.

**Quantitative results**

All cases were classified en two huge groups attending hormonal receptors immunoexpression into Invasive Ductal Carcinoma (IDC), and Invasive Lobular Carcinoma (ILC).

**Invasive Ductal Carcinoma (IDC):**

*Groups of patients without positive hormonal receptors*: In this group (n=7), 6 cases were Nm23-positive. The Nm23 index ranged between 2 and 97%, averaging 28.51%. LPA1 expression was found in 5 cases. LPA1 index ranged between 0 and 43%, averaging 8.15%. All cases were immunopositive to Ki67. The Ki67 averaging was 25.57% and Ki67 index ranged between 18 and 35%. p53 immunoexpression was observed in 2 cases, but index range were wide, between 0 and 95%, averaging 2.57%(Tables 2,3).

*Patient with ER positive*: Only one case was ER+ and PR-. All immunohistochemistry done were positive (see Tables 2,3). This group was not taken part in the study of correlations.

*Patient with PR positive*: Only one case was PR+ and ER-. As in the previous case, all markers were positive (Tables 2,3). This group neither formed part of the analysis of correlations.

*Groups of patients with positive hormonal receptors*: This group was the biggest (n=25), 23 cases were Nm23-positive. The Nm23 index ranged between 0 and 95%, averaging 42.08%. LPA1 expression was found in 17 cases. LPA1 index ranged between 0 and 63%, averaging 9.06%. All cases were immunopositive to Ki67. The Ki67 averaging was 15.64% and Ki67 index ranged between 10 and 25%. p53 immunoexpression was observed in 13 cases, index range between 0 and 95%, averaging 18%(Tables 2,3).

**Invasive Lobular Carcinoma (ILC):**

*Groups of patients without positive hormonal receptors*: No patients were included in this group (Tables 2,3).

*Patient with ER positive*: Only one case was ER+ and PR-. All immunohistochemistry done were positive (Tables 2,3). This group was not taken part in the study of correlations.

*Patient with PR positive*: No patients were included in this group (see Tables 2,3).

*Groups of patients with positive hormonal receptors*: In this group (n=15), 13 cases Nm23 positive, 13 cases LPA1 positive, 15 cases Ki67 positive, and 3 cases p53 positive were observed. The Nm23 index ranged between 0 and 94%, averaging 33.45%. LPA1 index ranged between 0 and 80%, averaging 13.97%. The Ki67 averaging was 11.2% and Ki67 index ranged between 7 and 13%. p53 index range between 0 and 2%, averaging 04%(Tables 2,3).

**Table 2.** Relationship between marker positivity and carcinomas classified according to hormonal receptors expression.

| Carcinoma | Group of patients | Ratio of positives cases | | | |
|---|---|---|---|---|---|
| | | Nm23 | LPA1 | Ki67 | p53 |
| IDC | ER- PR- | 6/7 | 5/7 | 7/7 | 2/7 |
| | ER+ PR- | 1/1 | 1/1 | 1/1 | 1/1 |
| | ER- PR+ | 1/1 | 1/1 | 1/1 | 1/1 |
| | ER+ PR+ | 22/25 | 17/25 | 25/25 | 13/25 |
| ILC | ER- PR- | 0/0 | 0/0 | 0/0 | 0/0 |
| | ER+ PR- | 1/1 | 1/1 | 1/1 | 1/1 |
| | ER- PR+ | 0/0 | 0/0 | 0/0 | 0/0 |
| | ER+ PR+ | 13/15 | 13/15 | 15/15 | 3/15 |

Invasive Ductal Carcinoma (IDC), and Invasive Lobular Carcinoma (ILC).

**Nm23 protein expression and its correlation with the clinicopathology variables**

Correlation between Nm23 and LPA1 and traditional prognostic markers in breast carcinoma was done, as well as Nm23/LPA1 and clinicopathology variables. Groups with 1 or 0 patients were eliminated. Only significant differences were found in Nm23/LPA1 and p53 in IDC ER- PR-. Pearson correlation coefficient is showed in Tables 4 and 5 and figures 5-7.

**DISCUSSION**

Several authors have been reported that ILC has characteristics and behaviors that are different from IDC [13;14], but it is unclear which of the two types showed better prognosis [15;16]. Therefore, in this study breast carcinoma samples had been separated into these two groups. After that, samples were classified attending the

**Table 3.** Relationship between marker positivity and carcinomas classified according to hormonal receptors expression.

| Carcinoma | Group of patients | Nm23 (%) | | LPA1 (%) | | Ki67 (%) | | p53 (%) | |
|---|---|---|---|---|---|---|---|---|---|
| | | Range | Mean | Range | Mean | Range | Mean | Range | Mean |
| IDC | ER- PR- | 2-97 | 28.51 | 0-43 | 8.15 | 18-35 | 25.57 | 0-95 | 2.57 |
| | ER+ PR- | 3 | - | 6 | - | 15 | - | 85 | - |
| | ER- PR+ | 75 | - | 16 | - | 10 | - | 5 | - |
| | ER+ PR+ | 0-95 | 42.08 | 0-63 | 9.06 | 10-25 | 15.64 | 0-95 | 18 |
| ILC | ER- PR- | - | - | - | - | - | - | - | - |
| | ER+ PR- | 1 | - | 3 | - | 30 | - | 30 | - |
| | ER- PR+ | - | - | - | - | - | - | - | - |
| | ER+ PR+ | 0-94 | 33.45 | 0-80 | 13.97 | 7-13 | 11.20 | 0-2 | 0.4 |

Invasive Ductal Carcinoma (IDC), and Invasive Lobular Carcinoma (ILC).

**Table 4.** Pearson correlation coefficients of Nm23 versus traditional prognostic markers in breast carcinoma.

| | IDC | | | | ILC | |
|---|---|---|---|---|---|---|
| | ER- PR- | | ER+ PR+ | | ER+ PR+ | |
| | r | Sig. | r | Sig. | r | Sig. |
| **LPA1** | -0,351 | p=0,441 | -0,2335 | p= 0,261 | -0,3624 | p= 0,184 |
| **Ki67** | -0,5185 | p=0,233 | -0,03895 | p= 0,853 | -0,06334 | p= 0,822 |
| **p53** | 0,7407 | p=0,057 | 0,2764 | p= 0,181 | 0,3810 | p= 0,161 |

r: Correlation, Sig.: significance; IDC: Invasive Ductal Carcinoma, and ILC: Invasive Lobular Carcinoma.

**Table 5.** Pearson correlation coefficients of Nm23/LPA1 versus traditional prognostic markers in breast carcinoma.

| | IDC | | | | ILC | |
|---|---|---|---|---|---|---|
| | ER- PR- | | ER+ PR+ | | ER+ PR+ | |
| | r | Sig. | r | Sig. | r | Sig. |
| **Ki67** | -0,4565 | p= 0,303 | -0,05506 | p= 0,798 | -0,05356 | p=0,849 |
| **p53** | 0,9707 | p<0.001 | -0,1242 | p= 0,563 | 0,3612 | p=0,186 |

r: Correlation, Sig.: significance; IDC: Invasive Ductal Carcinoma, and ILC: Invasive Lobular Carcinoma.

hormonal status of the patient. In 2005, the Committee of Consensus on Adjuvant Treatments for Breast Cancer at an Early Stage recommended that the first consideration in selecting the type of treatment should be the endocrine responsiveness. The recognition of this fact increased the relevance of pathologic evaluation as to the biological information [17].

For this reason the estrogen receptor has been the most extensively studied prognostic indicator to date. Orvieto et al. (2002) and Kasami et al. (2008) [18;19] have reported a positive relationship between estrogen receptor, increased disease-free interval and better survival of patients. The estrogen and progesterone receptors are currently the most widely used predictive factors for the choice of hormonal treatment [17].

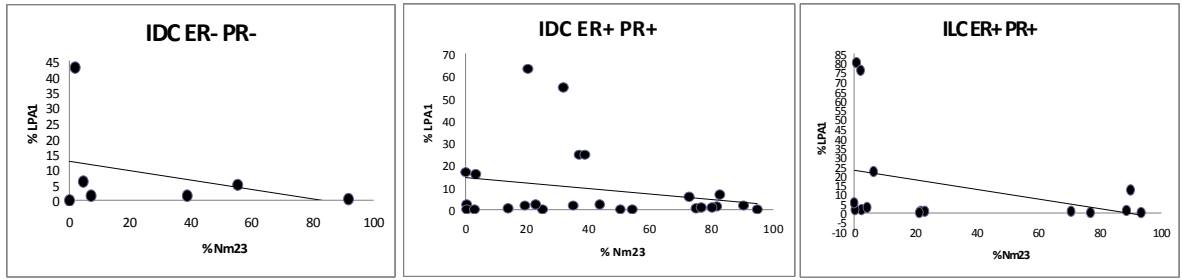

**Figure 5.** Dispersion graphs for %Nm23 and %LPA1. The three groups studied showed an inverse trend. IDC: Invasive Ductal Carcinoma; ILC: Invasive Lobular Carcinoma; ER: Estrogen Receptors, and PR: Progesterone Receptors.

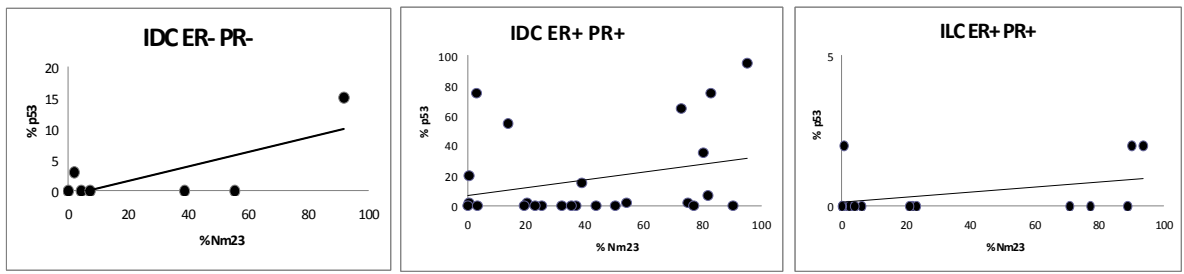

**Figure 6.** Dot plots for %Nm23 and %p53. All groups showed positive correlation, but not significant different was detected. IDC: Invasive Ductal Carcinoma; ILC: Invasive Lobular Carcinoma; ER: Estrogen Receptors, and PR: Progesterone Receptors.

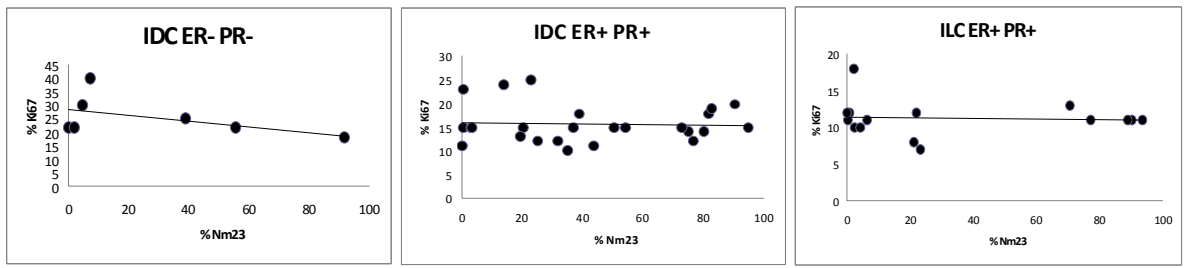

**Figure 7.** Dot plots for %Nm23 and %Ki67. Negative correlation was observed in studied groups. IDC: Invasive Ductal Carcinoma; ILC: Invasive Lobular Carcinoma; ER: Estrogen Receptors, and PR: Progesterone Receptors.

Antimetastatic genes, such as Nm23, play an important role in the porgression of breast cancer [20]. Several researchers have observed the probable inverse association of Nm23 expression with disease prognosis and/or metastasis [21]. Metastatic process involves activation and down regulation of multiple genes at each step of metastatic cascade [4]. Nm23 expression has been widely studied in various cancers and with their relation to staging and prognosis. Bal A et al. (2008) [4] results implicate that lack of Nm23 expression in early lesions may be predictive of progression to invasive carcinoma and thus could be helpful in predicting the aggressiveness of the disease.

In mestastatic cascade are implicated a number of molecules such as LPA1. Horak et al. (2007) [22] demostrated by *in vitro* studies, an inverse correlation between Nm23 and LPA1 . They observed *in vivo* that LPA1 could play an importan role in the promotion of metastasis. In our work, Pearson correlation coefficients Nm23 versus LPA1 were negative in all groups analysed. Not significant differences were detected probably because sample size was not enough big. However, the relationship between Nm23 and LPA1 showed an inverse correlation, as described Horak et al. (2007) [22].

p53 is the most frequently mutated gene in human cancer and is the founding member of the p53 family. In response to various cellular stresses, p53 regulates a variety of cellular functions including cell cycle progression, apoptosis, senescence, cell motility, DNA repair, genetic instability and cell metabolism by transcriptionally activating a variety of cellular genes [23-25]. Jung et al. (2007) [25] showed that Nm23-H1 and its binding partner STRAP (serine-threonine kinase receptor-associated protein) interacted with p53 and potentiated p53 activity. Both NM23-H1 and STRAP directly interacted with the central DNA binding domain within residues 113-290. The use of Nm23-H1 and STRAP mutants revealed that Cys(145) of Nm23-H1 and Cys(152) or Cys(270) of STRAP were responsible for p53 binding. Furthermore, Cys(176) and Cys(135) of p53 were required to bind Nm23-H1 and STRAP, respectively.

There is no consensus on expression of p53 when studying different carcinomas. Some authors associated p53 with a poor disease prognosis [26] and others with a high survival and good response to the treatment [27]. Simone et al. (2001) [28] found a positive correlation between presences of p53 with that of Nm23 histological samples, but Midulla el at. (1999) [29] were not observed any correlation within these two markers. This study is accorded with Jung et al. (2007) [25] and Simone et al. (2001) [28]. Note that the ratio Nm23/LPA1 was statistically significant for the group IDC ER- PR-. Maybe LPA1 plays an opposite role in p53 expression. In this regard, Hurst-Kennedy et al. (2009) [30] showed in an *in vitro* study with chondrocytes that LPA1 prevented apoptotic signalling by decreasing the abundance, nuclear localization, and transcriptional activity of the tumour-suppressor p53.

Ki67 is expressed in G1-, S-, G2 and mitosis phase. Nm23 has been correlation inversely with Ki67 in melanoma of the oral cavity and bladder tumour, showing a low expression of Ki67 in benign tissues and high expression of Nm23 in this tumour type [31]. Also, decrease of Nm23 expression, but not Ki-67, was significantly correlated with lymph node metastasis of breast invasive ductal carcinoma [32]. In all cases studied, there is an inverse correlation between Nm23 or Nm23/LPA1 and Ki67, although none of them was significant.

In summary, it can be concluded that: 1) Nm23 and LPA1 tend to be inversely correlated; 2) Nm23 correlates positively with p53, regardless of tumour type and patient's hormonal status; 3) Nm23 shows a reverse trend in comparison with Ki67. However, more studies are needed to determinate if hormone receptors play some role in the inverse relationship between Nm23-LPA1 and to know the tumour type implication in these relationships.

## ACKNOWLEDGEMENTS

We are grateful to R. Romero Osuna and F. Vidal PhD (Laboratory of Histology and Imaging, Institute of Applied Molecular Medicine Institute, CEU San Pablo University, School of Medicine, Madrid, Spain), for their expert assistance with histological techniques.

This study was supported by financial grant from San Pablo-CEU University and Santander Bank (USP-BS-PPC04/2011).

## REFERENCE

[1] Panupinthu N, Lee HY, Mills GB. Lysophosphatidic acid production and action: critical new players in breast cancer initiation and progression. Br J Cancer 2010; 102: 941-946.

[2] Weigelt B, Geyer FC, Reis-Filho JS. Histological types of breast cancer: how special are they? Mol Oncol 2010; 4: 192-208.

[3] Castellana B, Escuin D, Peiro G, Garcia-Valdecasas B, Vazquez T, Pons C, Perez-Olabarria M, Barnadas A, Lerma E. ASPN and GJB2 Are Implicated in the

Mechanisms of Invasion of Ductal Breast Carcinomas. J Cancer 2012; 3: 175-183.

[4] Bal A, Joshi K, Logasundaram R, Radotra BD, Singh R. Expression of nm23 in the spectrum of pre-invasive, invasive and metastatic breast lesions. Diagn Pathol 2008; 30: 23.

[5] Stoll BA. Premalignant breast lesions: role for biological markers in predicting progression to cancer. Eur J Cancer 1999; 35: 693-697.

[6] Steeg PS, Bevilacqua G, Kopper L, Thorgeirsson UP, Talmadge JE, Liotta LA, Sobel ME. Evidence for a novel gene associated with low tumor metastatic potential. J Natl Cancer Inst 1988; 80: 200-204.

[7] Steeg PS. Search for metastasis suppressor genes. Invasion Metastasis 1989; 9: 351-359.

[8] Steeg PS, Horak CE, Miller KD. Clinical-translational approaches to the Nm23-H1 metastasis suppressor. Clin Cancer Res 2008; 14: 5006-5012.

[9] Takuwa Y, Takuwa N, Sugimoto N. The Edg family G protein-coupled receptors for lysophospholipids: their signaling properties and biological activities. J Biochem 2002; 131: 767-771.

[10] Moolenaar WH, van Meeteren LA, Giepmans BN. The ins and outs of lysophatidic acid signaling. Bioessays 2004; 26: 870-881.

[11] Kohlberger PD, Breitenecker F, Kaider A, Losch A, Gitsch G, Breitenecker G, Kieback DG. Modified true-color computer-assisted image analysis versus subjective scoring of estrogen receptor expression in breast cancer: a comparison. Anticancer Res 1999; 19: 2189-2193.

[12] Maximova OA, Taffs RE, Pomeroy KL, Piccardo P, Asher DM. Computerized morphometric analysis of pathological prion protein deposition in scrapie-infected hamster brain. J Histochem Cytochem 2006; 54: 97-107.

[13] Chia S, Al-Foheidi M, Speers C, Woods R, Kennecke H. 217 Outcome of invasive lobular carcinoma compared to infiltrating ductal carcinoma: a population based study from British Columbia. EJC Supplements 2010; 8: 122.

[14] Lee JH, Park S, Park HS, Park BW. Clinicopathological features of infiltrating lobular carcinomas comparing with infiltrating ductal carcinomas: a case control study. World J Surg Oncol 2010; 8: 34.

[15] Cristofanilli M, Hayes DF, Budd GT, Ellis MJ, Stopeck A, Reuben JM, Doyle GV, Matera J, Allard WJ, Miller MC, Fritsche HA, Hortobagyi GN, Terstappen LW. Circulating tumor cells: a novel prognostic factor for newly diagnosed metastatic breast cancer. J Clin Oncol 2005; 23: 1420-1430.

[16] Pestalozzi BC, Zahrieh D, Mallon E, Gusterson BA, Price KN, Gelber RD, Holmberg SB, Lindtner J, Snyder R, Thurlimann B, Murray E, Viale G, Castiglione-Gertsch M, Coates AS, Goldhirsch A. Distinct clinical and prognostic features of infiltrating lobular carcinoma of the breast: combined results of 15 International Breast Cancer Study Group clinical trials. J Clin Oncol 2008; 26: 3006-3014.

[17] Pachnicki JP, Czeczko NG, Tuon F, Cavalcanti TS, Malafaia AB, Tuleski AM. Immunohistochemical evaluation of estrogen and progesterone receptors of pre and post-neoadjuvant chemotherapy for breast cancer. Rev Col Bras Cir 2012; 39: 86-92.

[18] Orvieto E, Viale G. Receptores dos hormônios esteroides. In. Rio de Janeiro:Medsi: 2002: 267-271.

[19] Kasami M, Uematsu T, Honda M, Yabuzaki T, Sanuki J, Uchida Y, Sugimura H. Comparison of estrogen receptor, progesterone receptor and Her-2 status in breast cancer pre- and post-neoadjuvant chemotherapy. Breast 2008; 17: 523-527.

[20] Yamashita H, Kobayashi S, Iwase H, Itoh Y, Kuzushima T, Iwata H, Itoh K, Naito A, Yamashita T, Masaoka A. Analysis of oncogenes and tumor suppressor genes in human breast cancer. Jpn J Cancer Res 1993; 84: 871-878.

[21] Tokunaga Y, Urano T, Furukawa K, Kondo H, Kanematsu T, Shiku H. Reduced expression of nm23-H1, but not of nm23-H2, is concordant with the frequency of lymph-node metastasis of human breast cancer. Int. J Cancer. 1993; %19;55: 66-71.

[22] Horak CE, Lee JH, Elkahloun AG, Boissan M, Dumont S, Maga TK, Arnaud-Dabernat S, Palmieri D, Stetler-Stevenson WG, Lacombe ML, Meltzer PS, Steeg PS. Nm23-H1 suppresses tumor cell motility by down-regulating the lysophosphatidic acid receptor EDG2. Cancer Res 2007; 67: 7238-7246.

[23] Chen YK, Huse SS, Lin LM. Differential expression of p53, p63 and p73 proteins in human buccal squamous-cell carcinomas. Clin Otolaryngol Allied Sci 2003; 28: 451-455.

[24] Lai D, Visser-Grieve S, Yang X. Tumour suppressor genes in chemotherapeutic drug response. Biosci Rep 2012; 32: 361-374.

[25] Jung H, Seong HA, Ha H. NM23-H1 tumor suppressor and its interacting partner STRAP activate p53 function. J Biol Chem 2007; 282: 35293-35307.

[26] Lee KE, Lee HJ, Kim YH, Yu HJ, Yang HK, Kim WH, Lee KU, Choe KJ, Kim JP. Prognostic significance of p53, nm23, PCNA and c-erbB-2 in gastric cancer. Jpn J Clin Oncol 2003; 33: 173-179.

[27] Shun CT, Wu MS, Lin JT, Chen SY, Wang HP, Lee WJ, Wang TH, Chuang SM. Relationship of p53 and c-erbB-2 expression to histopathological features, Helicobacter pylori infection and prognosis in gastric cancer. Hepatogastroenterology. 1997; 44: 604-609.

[28] Simone G, Falco G, Caponio MA, Campobasso C, De Frenza M, Petroni S, Wiesel S, Leone A. nm23 expression in malignant ascitic effusions of serous ovarian adenocarcinoma. Int J Oncol 2001; 19: 885-890.

[29] Midulla C, De Iorio P, Nagar C, Pisani T, Cenci M, Valli C, Nofroni I, Vecchione A. Immunohistochemical expression of p53, nm23-HI, Ki67 and DNA ploidy: correlation with lymph node status and other clinical pathologic parameters in breast cancer. Anticancer Res 1999; 19: 4033-4037.

[30] Hurst-Kennedy J, Boyan BD, Schwartz Z. Lysophosphatidic acid signaling promotes proliferation, differentiation, and cell survival in rat growth plate chondrocytes. Biochim Biophys Acta 2009; 1793: 836-846.

[31] Korabiowska M, Honig JF, Jawien J, Knapik J, Stachura J, Cordon-Cardo C, Fischer G. Relationship of nm23 expression to proliferation and prognosis in malignant melanomas of the oral cavity. In Vivo 2005; 19: 1093-1096.

[32] Terasaki-Fukuzawa Y, Kijima H, Suto A, Takeshita T, Iezumi K, Sato S, Yoshida H, Sato T, Shimbori M, Shiina Y. Decreased nm23 expression, but not Ki-67 labeling index, is significantly correlated with lymph node metastasis of breast invasive ductal carcinoma. Int J Mol Med 2002; 9: 25-29.